\documentstyle[12pt,aps]{revtex}
\begin{document}
\title{ Annealing of defects in Fe after MeV Heavy ion irradiation} 
\author{G. Aggarwal and P. Sen}
\address{School of Physical Sciences, Jawaharlal Nehru University,
New Delhi-110067, India}
\maketitle
\vskip .5cm
\noindent
\abstract
{We report study of recovery dynamics, followed by in-situ resistivity
measurement after 100 MeV oxygen ion irradiation, in cold rolled Fe at
300K. Scaling behavior with microstructural density and temperature of sample
have been used to establish stress induced defects formed during irradiation 
as a new type of sink.
The dynamics after irradiation has been shown to be due to migration of
defects to two types of sinks i.e. stress induced defect as variable sinks
and internal surfaces as fixed sinks. Experimental data obtained under
various experimental conditions have been fitted
to theoretical curves. Parameters thus obtained from fitting are employed
to establish effect of electronic energy loss and temperature on
recovery dynamics and stress associated with variable sinks.}

\newpage
Transfer of energy after inelastic collision suffered by swift
MeV ions, to lattice atoms, is well established now. The
dissipation of ion energy takes place in two stages. The lattice
electrons initially pick up large kinetic energies which couple
to the lattice atoms through electron - phonon (e-p) coupling.
This is seen to induce defect production in several metals (Fe,
Ni, Ti etc.)~\cite{ai,hd1}, where e-p coupling is large, and
also in metals where this is not very large such as Cu and
Ag~\cite{ai1}. Moreover, in metals where very strong e-p
coupling exist (Al, Ni and Pt)~\cite{ai,ai2} an anomalous
reduction in defect density is noticed. This is ascribed to
radiation annealing, leading to decay of single interstitials.

Several years ago, quenching-in of vacancies in noble metals was
established by Bauerle and Koehler~\cite{bk}, Kimura et
al.~\cite{k} and De Jong and Koehler~\cite{md} by following
their annealing behavior employing electrical resistivity
measurements. These were later substantiated by transmission
electron microscopy (TEM)~\cite{sh,c}.  The relative decrease in
quenched-in resistivity as a function of time showed S-shaped
curves instead of an exponential decay expected due to
annihilation of single vacancies with high activation energies
of motion.  Departure from exponential decay was assigned to
formation of divacancies, vacancy clusters and deposition of
vacancies at sessile ring sites with low activation energy.

Swift MeV ions, passing through thin specimens, subject the
material predominatly to electronic excitations as the ion
ranges are sufficiently large. Following e-p coupling, the
lattice atoms are agitated around the ion irradiation site,
possibly in concentric circles (assuming electronic conductivity
to be isotropic for a polycrystalline specimen). Under low flux
these circles will not overlap. This will lead to regions which
are at high temperature directly in contact with cold regions,
contiguous in space. A situation similar to quenching can arise
if the hot-cold interface temperature difference is reasonably
large, allowing the formation of vacancies. This will probably
be most efficient at natural interfaces, for example, at
dislocation and grain boundaries of cold rolled metallic foils.

In this Letter we shall try to establish result of electronic
excitation in a pure Fe foils by following time evolution of
electrical resistivity after subjecting them to MeV ions. This
method not only allows us to confirm defect production through
MeV ion irradiation but shed futher light on the long term
annealing products such as vacancy clusters or sessile rings,
as established before employing other procedures~[5-7].

Polycrystalline metal foils (99.99\% pure, Goodfellow, UK) were
irradiated with 100 MeV oxygen ions ($S_e \sim 2 MeV/\mu m$)
accelerated with the help of a 15 UD pelletron at NSC, New
Delhi. A constant flux of $8.8 \times
10^{9}~ions~cm^{-2}~sec^{-1}$ was maintained normal to the
surface of the sample while an in-situ four probe resistivity
measurement at constant sampling current monitored the time
evolution of defect density. Response Voltage (RV) was collected
every t= 0.2 sec using a computer controlled Keithley nano
voltmeter. Foils (10$\mu m$ thick) of Fe were mounted on mica or
sapphire substrates for electrical isolation from the sample
holder. Temperature during irradiation, monitored with the help
of a thermocouple placed next to the sample, was held constant
by thermalising with a large constant flow copper cryostat. The
low ion flux maintained on the sample does not produce any
temperature rise. This is confirmed by directly measuring on
thermocouple. The MeV ions on entering solid targets lose part
of their energy to the nuclear ($S_n$) and electronic ($S_e$)
subsystems. The two loss processes can be spatially separated by
appropriately choosing the thickness of the target. In this
study, changes in the intrinsic physical property of the system
would be primarily $S_e$ related as ion range (from
TRIM~\cite{nh} code is $37\mu m$) is much larger than the thickness
of Fe foils.

In an earlier report we~\cite{ps} showed evidence of strain
induced by electronic energy loss using x-ray topography (XRT)
for strain mapping in a Si(001) single crystal. In
polycrystalline metal foils, individual grains can be thought of
as ``single crystals'' with dislocation lines providing natural
interfaces. Under stress, generated by ion irradiation at an
interface, atoms in a dislocation line can be moved to generate
line defects termed as stress induced defects (SID). Motion of
atoms at dislocations generated this way is reversible as the
stress induced by $S_e$, when removed, initiates a fall of
resistivity to its pre-irradiated value. The reversible motion
of dislocations under stress, contributes a reversible plastic
strain component to the total strain, and can be detected during
measurement of strain dependent physical parameters.  These
defects are different from point defects as has been established
by their different scaling behavior with temperature~\cite{ps}.
Under favorable conditions, positive feedback sets in, resulting
in formation of dissipative structures~\cite{ss}. These
structures are maintained as long as energy is supplied by the
incoming ions. On switching the beam off, these structures decay
along with decay of SID's resulting in jumps in RV. 
This decay is followed by migration of
defects to remnant structures left after the dissociation of
dissipative structures at internal surfaces.  

In this experiment we try to understand in detail the dynamics
involved during defect migration far away from the irradiation
event. In Fig.1 we show plot of fraction of defects remaining
($\frac{\Delta RV}{\Delta RV_o}$) vs time (t) for cold rolled Fe
after 100 MeV oxygen ion beam was switched off at room
temperature. The data point at t~=~0 delineates the SID region
from the annealing region and hence signifies starting of
annealing behavior. Fraction of defects remaining is defined as 
$$ \frac{\Delta RV}{\Delta RV_o} = \frac{RV(t) - RV(t=
infinity)}{RV(t=0) -RV(t=infinity)}$$ 
Polycrystalline metals are characterised by large
amount of inherent defect structures like grain boundaries,
dislocation lines etc. These structures are efficient sinks for
non-equilibrium defects. Under the influence of thermodynamic
forces non-equilibrium defects migrate to sinks.  Assuming the
number of sinks to remain constant in time, annealing of defects
takes place exponentially~\cite{d}. Under certain thermodynamic
conditions a new type of sink is formed, identified in the
literature as a variable sink. This sink is charaterised by
explicit time dependence on the number of sinks. Annealing of
defects to variable sinks generate S-shaped decay curves.
Presence of both type of sinks in any material would generate
decay curves having the characteristics of fixed and variable
sink decay. Depending on relative density of each type of sinks,
two cases are important :(1) density of variable sinks is larger
than density of fixed sinks and (2) density of variable sinks is
comparable to density of fixed sinks. In the former case functional
form of decrease in defect concentration with time is given
as~\cite{k}
\begin{equation}
V(t)  = \frac {V_o}{(Cosh(\beta t))^2}
\end{equation}
where,$ V_o$ is the concentration of defects before annealing
and, $\beta $ is the rate parameter.  In the latter case, the
decay process is a combination of first order kinetics due to
migration of defects to fixed sinks and decay originating due to
variable sinks. Assuming the number of defects migrating to
fixed sinks to be independent of concentration of variable sinks
and vice versa, decay is solution of two simultaneous
differential equations~\cite{d}. Analysis of the above decay is
too involved to allow determination of physical parameters from
our experimental curves due to lack of data on concentration of
different types of sinks. However an analysis based on fitting
procedure is possible by considering the decay to be composed of
additive parts as described above. Functional form of the decay
would be
\begin{equation}
V(t)  = \frac {V_o}{(Cosh(\beta t))^2} + A e^{-t/{\tau}}
\end{equation}

Good fitting was obtained for decay dynamics in Fig.1 with equation 2.
Fitting to the experimental data was helped by the following
observations: (1) tail of 
decay curve is dominated by the exponential function and provides the slope of
exponential fall and (2) initial decay is dominated by variable sink and
hence would provide an estimate of $\beta$.
The solid line represents a theoretical fit to the experimental data
(circle). Curves a and b represent the individual contribution of variable
sink and fixed sink to total recovery dynamics. Ratio of $V_o$
and $A$, $R_{V_o/A}$ is a quantity signifying contribution of variable
sink with respect to fixed sink. Values obtained for the quantities
$R_{V_o/A}$, $\beta$ and $\frac{1}{\tau}$ from fitting are 23.89,
0.136633~$\pm$~0.0005 $sec^{-1}$ and 0.059~$\pm$~0.014
$sec^{-1}$ respectively. 

It is well known that point defects exist in metals at all temperatures
but the concentration is in thermal equilibrium and hence
low. In order to study properties of defects like
energy of migration, energy of formation etc. and their effects on
physical properties of a host metal, defects are engineered in excess
or non-equilibrium concentrations. The process of allowing
non-equilibrium concentration of defects at a given temperature to come to
equilibrium is termed as annealing. Recent investigation of
Ti~\cite{hd1,hd} under high energy ion irradiation showed phase 
transformation from $\alpha$ phase to metastable $\omega$ phase. It is
known that Ti undergoes this kind of transformation when
subjected to high static
pressure of 8 Gpa~\cite{v,s} for about 24 hours. Similarities observed in
effects due to irradition and pressure supports us in using kinetics
observed under quenching to kinetics observed under irradiation. In our
experiment we engineer non-equilibrium concentration of defects using high
energy ion irradiation. Behavior of RV after switching the beam off would fall in the regime of
annealing theory and plots obtained would be termed as annealing plots.

In order to justify SID as variable sinks we
compare annealing plots of Fe foils after 200 MeV silver ion irradiation for
different temperatures and microstructural density. In Fig.2a we show
$\frac{\Delta RV}{\Delta RV_o}$ vs t for cold rolled
and annealed Fe.  Annealed Fe shows an exponential decay
after an initial rise whereas cold rolled Fe shows an S-shaped decay
charateristic of variable sinks. Under high energy ion irradiation
individual ions deposit large 
amount of energy in a local region. This region thermalises in time scales
of the order of $10^{-12}~sec$ thus simulating conditions similar to
quenching. Variable sinks have been observed in quenched gold, with divergent 
models being proposed of their geometry and process of production~[5-9]. Under 
irradiation, if variable sinks originate due to quenching then annealed Fe
should also show S-shaped decay. But this is certainly not
observed. We have shown already established of SID in annealed Fe~\cite{ps}.
Another aspect is the time required to fall to half its initial 
value i.e. $t_{\frac{1}{2}}$,
which defines the sharpness of the decay. This is less in the case of 80K 
annealing than at 300K. Finally, in Fig.2b we
show plots of $\frac{\Delta RV}{\Delta RV_o}$ vs t in 
cold rolled Fe at two different irradiating and annealing temperatures 
of 300K and 80K. Observation of S-shaped
decay curves at 300K could be attributed to migration of defects to remnant
structures left after release of dissipative structures or to low
concentration of SID. Similar decay curves are seen in absence of
dissipative structure formation at 80K. This together with
absence of S-shaped decay in annealed Fe conclusively proves role of
SID as variable sinks. 

After having established the contribution of SID towards
variable sinks we proceed to analyse the effect of $S_e$ on
recovery dynamics.  In Fig.3a we show annealing plot of
$\frac{\Delta RV}{\Delta RV_o}$ vs t for cold rolled Fe after
200 MeV silver ion irradiation ($S_e \sim 15 MeV/\mu m$) at
300K. Samples were irradiated for 30 minutes before measurements
were taken. Experimental data (circle) fitted well to pure
variable sink decay dynamics given by equation 1 (solid line).
Values obtained for different parameters $V_o$ and $\beta$ from
fitting are 1.0 and 0.116~$\pm$~0.00049 $sec^{-1}$ respectively. Change in
functional form to pure variable sink decay after silver
irradiation, from a combination decay after oxygen irradiation,
could be attributed to difference in flux or to difference in
$S_e$ values. In Fig.3b we show annealing plots of $\frac{\Delta
RV}{\Delta RV_o}$ vs t for cold rolled Fe at two ion currents of
1.3 nA and 1.8 nA. Decrease in $t_{\frac{1}{2}}$ is observed for
increase in ion flux. Theoretical fitting to the data showed
increase in value of $\beta$ with decrease in $t_{\frac{1}{2}}$.
Increase in value of $\beta$ has been attributed in the
literature to increase in concentration of variable
sink~\cite{d}. Hence flux has the effect of increasing the
concentration of variable sink but not to change the functional
form of decay. It is also observed that same value of
$\beta$~=~0.136633 $sec^{-1}$ is obtained in silver irradiation with one
order of lower flux than oxygen irradiation.  Silver ions
dissipate higher $S_e$ and present larger stress across the
boundaries thus creating a large fraction of SID. These SID's
with a range of activation energies
are formed across dislocation lines, reducing the fraction of
fixed sinks. This shows that ions with higher $S_e$ transfer
larger energy to the lattice and are more
efficient in creating SID. This also explains pure variable sink
recovery dynamics under silver irradiation.

Variation of recovery dynamics with temperature has been
established in the literature~\cite{d} is due to  
(1) nature of defects created and (2) mobility of defects.
The former would affect parameters $R_{V_o/A}$ and $\beta$ and latter
$\frac{1}{\tau}$. In
Fig.4 we show $\frac{\Delta RV}{\Delta RV_o}$ vs t for 
cold rolled Fe under silver irradiation, at 80K. Good agreement with the
experimental data was found employing equation 2. A solid line shows the
theoretical fit to experimental data (circles). Individual contribution of 
variable and fixed sinks are represented by curves a and b respectively.
Fitting gives the values of
$R_{V_o/A}$, $\beta$ and $\frac{1}{\tau}$ as 2.0,
0.1905~$\pm$0.002 $sec^{-1}$ and 0.0837~$\pm$0.003 $sec^{-1}$
respectively.  On comparing
with recovery dynamics at 300K (Fig.3b) the following
observations are recorded: (1) change in functional form of recovery
dynamics and (2) increase in value of $\beta$. 
At lower temperatures grain become weaker than grain boundaries,
making it difficult to create SID at internal surfaces. 
Thus we have decrease in concentration of variable sinks with respect
to its concentration at 300K. The decrease in 
variable sinks increases the fraction of fixed sinks and results in 
combination decay with a lower value of $R_{V_o/A}$. Parameter $\beta$ is
a function of the concentration of variable sink, rate
constant ($\frac{1}{\tau}$) and concentration of conventional point
defects. Decrease in rate constant with decrease in temperature is well
known in published literature~\cite{d}. Increase in $\beta$ at
lower temperature is a result of 
substantial increase in concentration of conventional point defects.

In annealing theory sinks are modeled without a stress field.
Relaxation time ($\tau$) is known to be influenced by the stress
field of sinks and temperature of annealing~\cite{d}. In order
to establish that stress fields are associated with SID we study
annealed Fe after irradiation. Figure 5 shows
$\frac{\Delta RV}{\Delta RV_o}$ vs t after the
beam has been switched off. It is assumed that in samples with low
SID concentration, annealing starts immediately
after switching the beam off. In Fig.5 we see an initial rise
which could posssibly be due to breakdown of irradiation induced
clusters. This rise has been neglected for the moment
and the decay part fitted to the desired function. An
exponential function is in 
excellent agreement with the experimental data (circle). Fitting
gives the value of $\frac{1}{\tau}$ as 0.0953 $sec^{-1}$.
Dislocation lines are source of variable sink and in annealed Fe
low density of dislocations result in low density of variable
sinks. In presence of an insignificant density of variable sink,
the decay dynamics is exponential in form. The value
of $\tau$ obtained from the fitting is 10 sec. This value is in
absence of stress field associated with SID. Assuming that
decrease in $S_e$ effects the ratio of variable sink to fixed
sink only, value of $\tau$ in presence of SID is 20 sec
(Fig.3a). Such a large increase in $\tau$ for same sample
temperature can be only associated to stress field of SID.

Although we have shown recovery following quenching and heavy
ion irradiation to have similar overall behavior, their dynamics
differ on two accounts. Firstly, under irradiation cold rolled
Fe showed a S-shaped recovery curve which on annealing show
exponential decay (Fig.2a).  Whereas in the quenching experiment
reported in the literature~[5-7], annealed gold shows S-shaped
recovery curve while cold rolled gold wire shows exponential
decay. Secondly, under irradiation the time scale of recovery is
in order of seconds whereas in quenching it is of the order of
hours. The former difference is due to different nature of
variable sinks created under the two processes while for the
latter, difference in time scales can be attributed to the
nature of stress field associated with SID.

In conclusion, we have suggested that new stress induced defects
(SID) behave like variable sinks following heavy ion
irradiation. Recovery of defects was a result of their migration
to two types of sinks, fixed and variable. It was also found to
be sensitive to $S_e$ and temperature of the sample. Stress
associated with SID have been shown to effect the relaxation
time of defect migration to sinks.

We acknowledge the cooperation extended by NSC, New Delhi during
this study. Partial financial support from UGC is also acknowledged. 

\newpage
\centerline{\bf Figure Captions}
\begin{itemize}
\item Fig.1~: Recovery curve in cold rolled Fe followed by in-situ
resistivity measurement after 100 MeV oxygen ion irradiation at 300K.
Experimental data (circle) is found to have S-shaped decay
charateristic of variable sink. Solid curves a and b represent the
contribution of variable sink and fixed sink to total decay.
Time (t=0) here and in all subsequent plots for cold rolled (300K)
corresponds to starting of annealing behaviour after SID decay~[11]
\item Fig.2~: Comparative recovery curves for (a)
microstructural density and (b) temperature of recovery, after 200 MeV silver
ion irradiation in thin Fe foils. These curves have been
analysed in detail in subsequent figures. Due to absence of SID
time (t=0) for annealed Fe (a) and cold rolled Fe at 80K (b)
corresponds to beam switch off
\item Fig.3~: Recovery curves in cold rolled Fe after 200 MeV silver ion
irradiation at 300K, where,(a) decay is found to be controlled by pure variable
sinks and (b) functional form of decay is unaffected by ion flux
\item Fig.4~: Recovery curves in cold rolled Fe after silver irradiation at
80K. Contribution of variable sink (curve a) decreases and that of fixed
sink (curve b) increases when compared to recovery at 300K. Ion
current was 1.5nA 
\item Fig.5~: Recovery curve in annealed Fe after silver irradiation at
300K. Solid line is the exponential fit to experimental data
data (circle). The ion current was 3.0nA
\end{itemize}
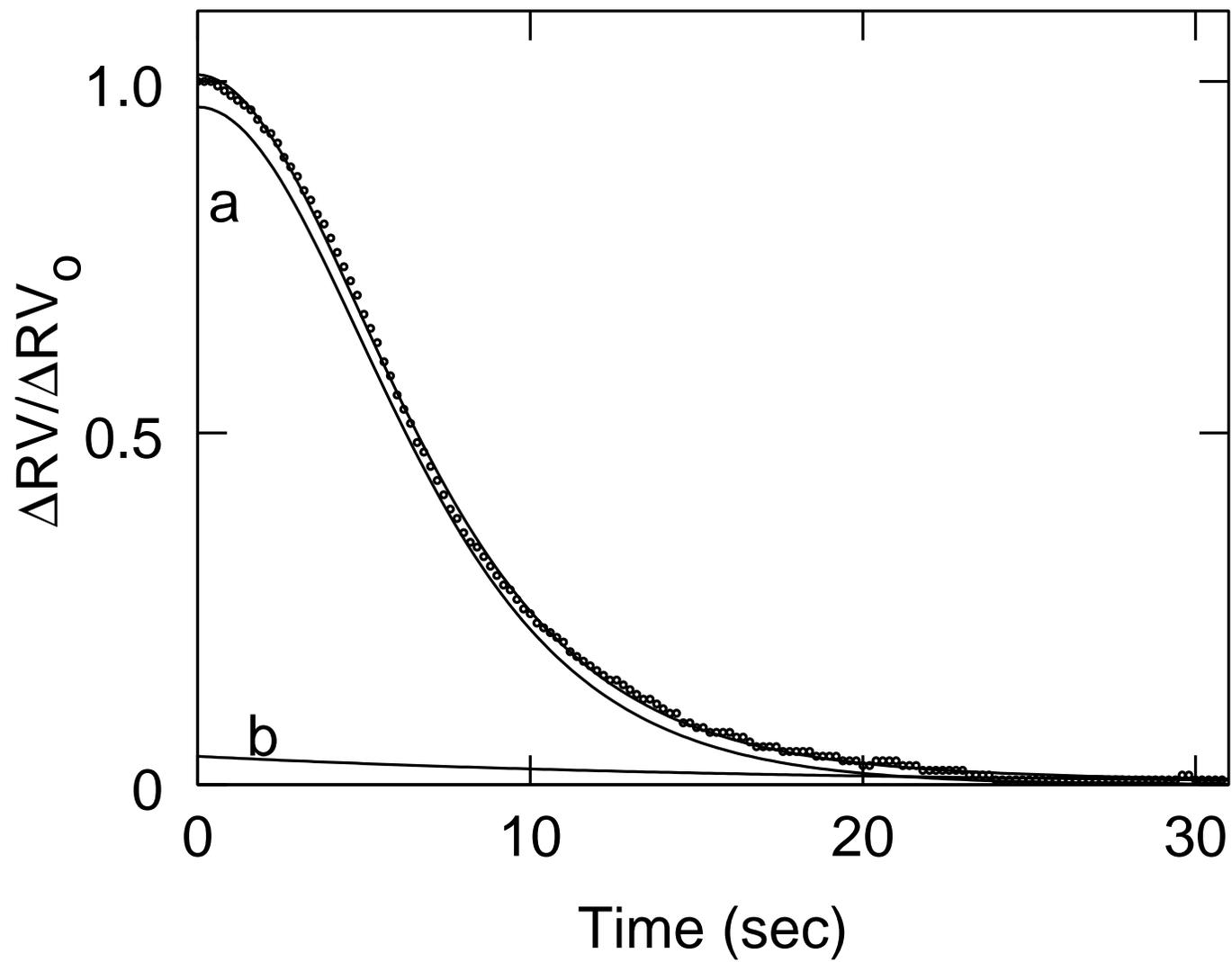
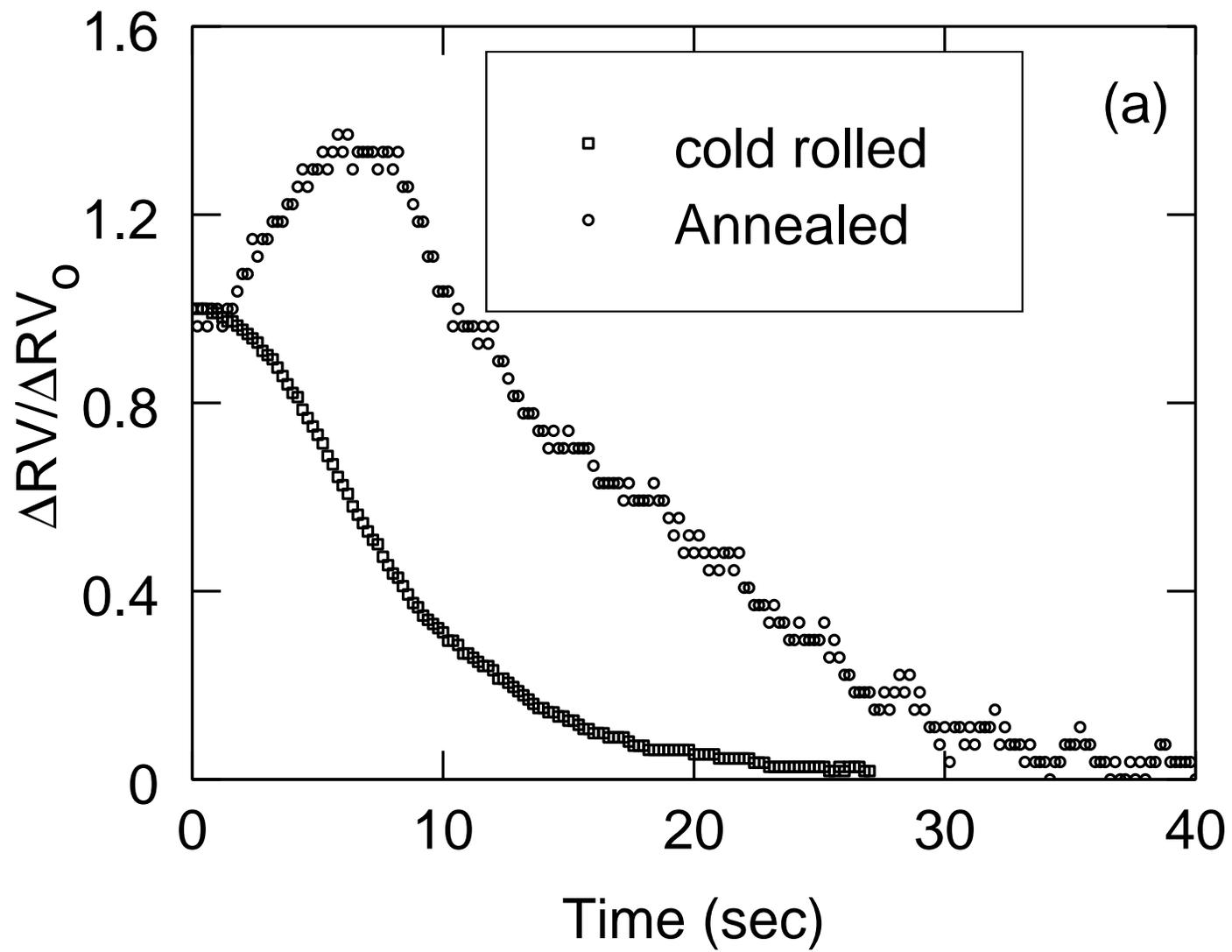
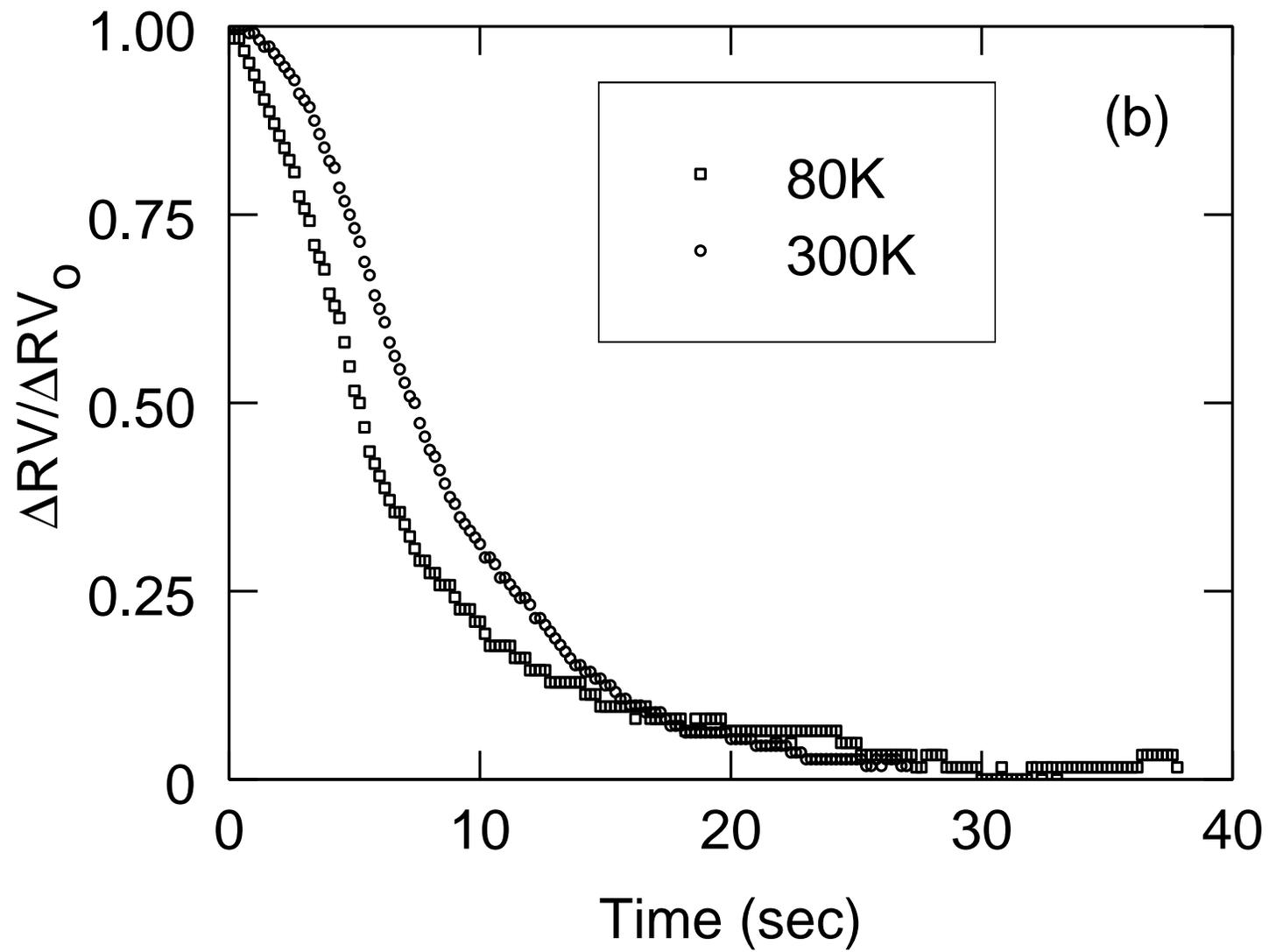
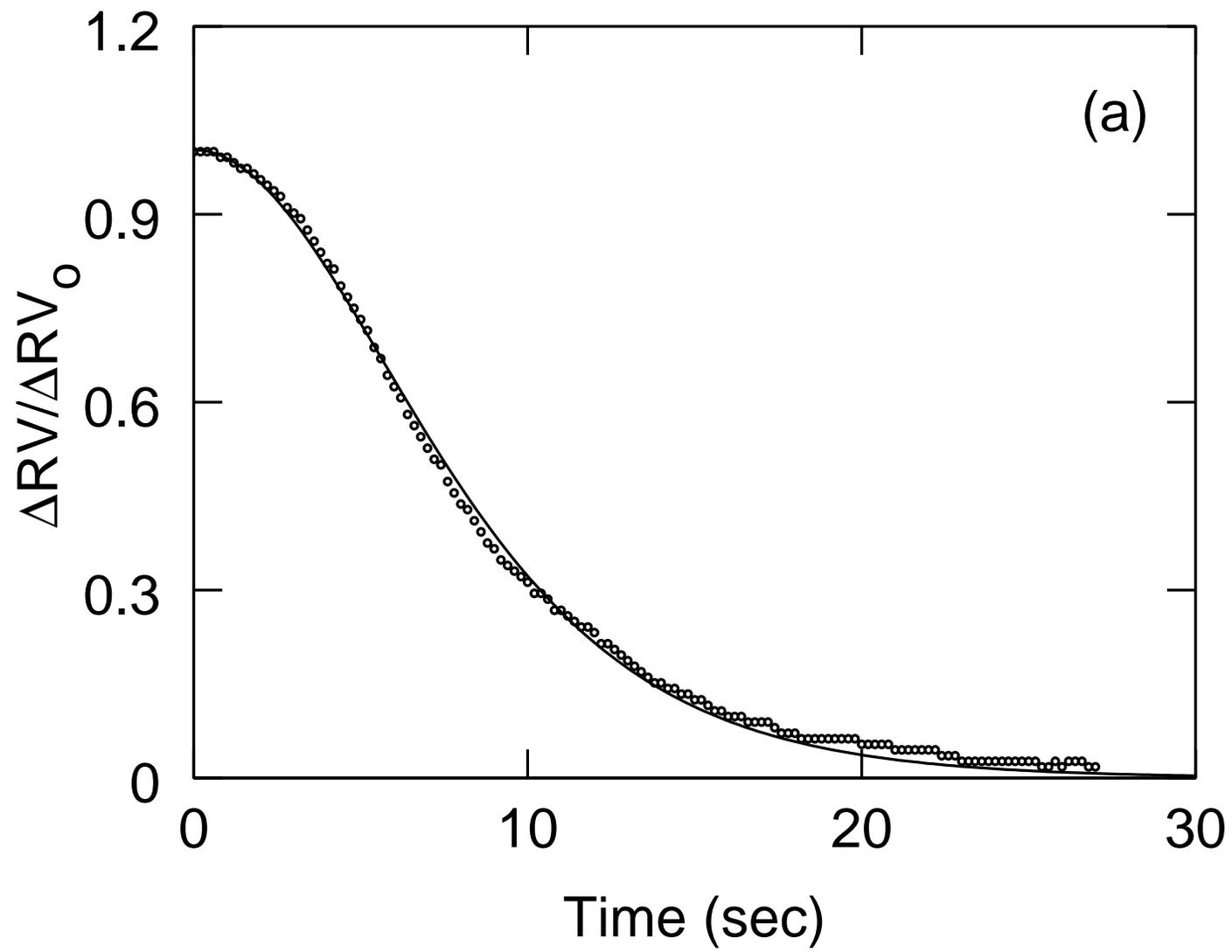
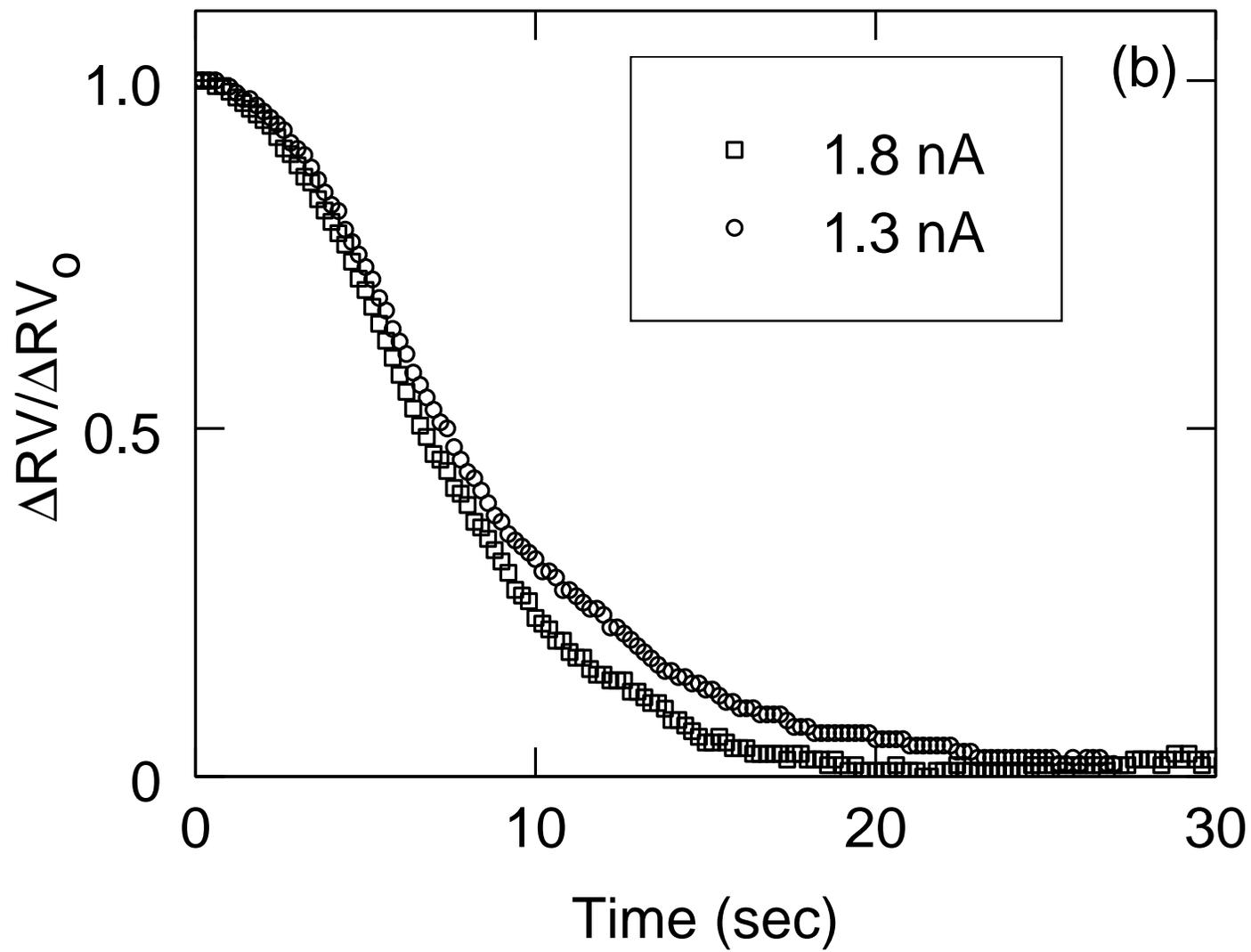
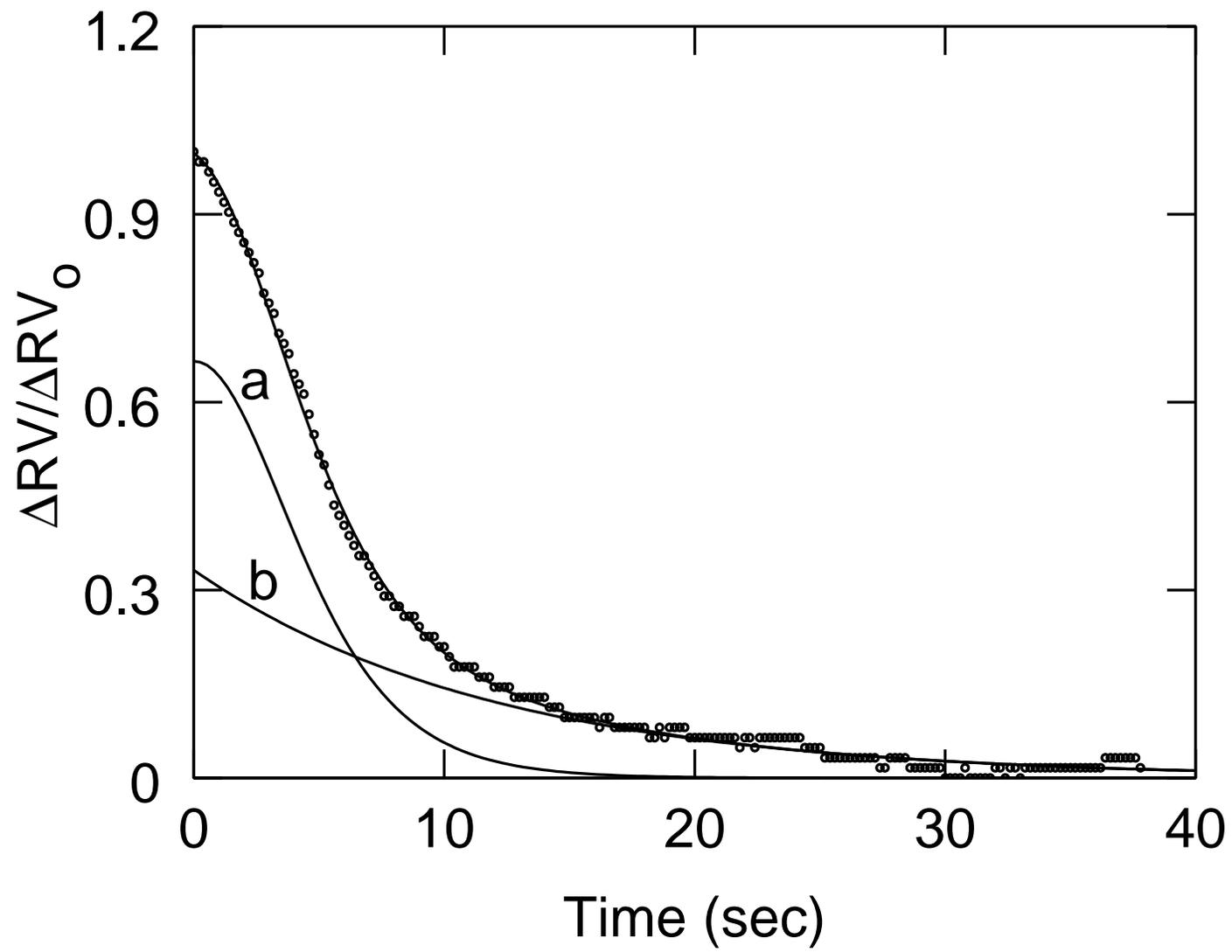
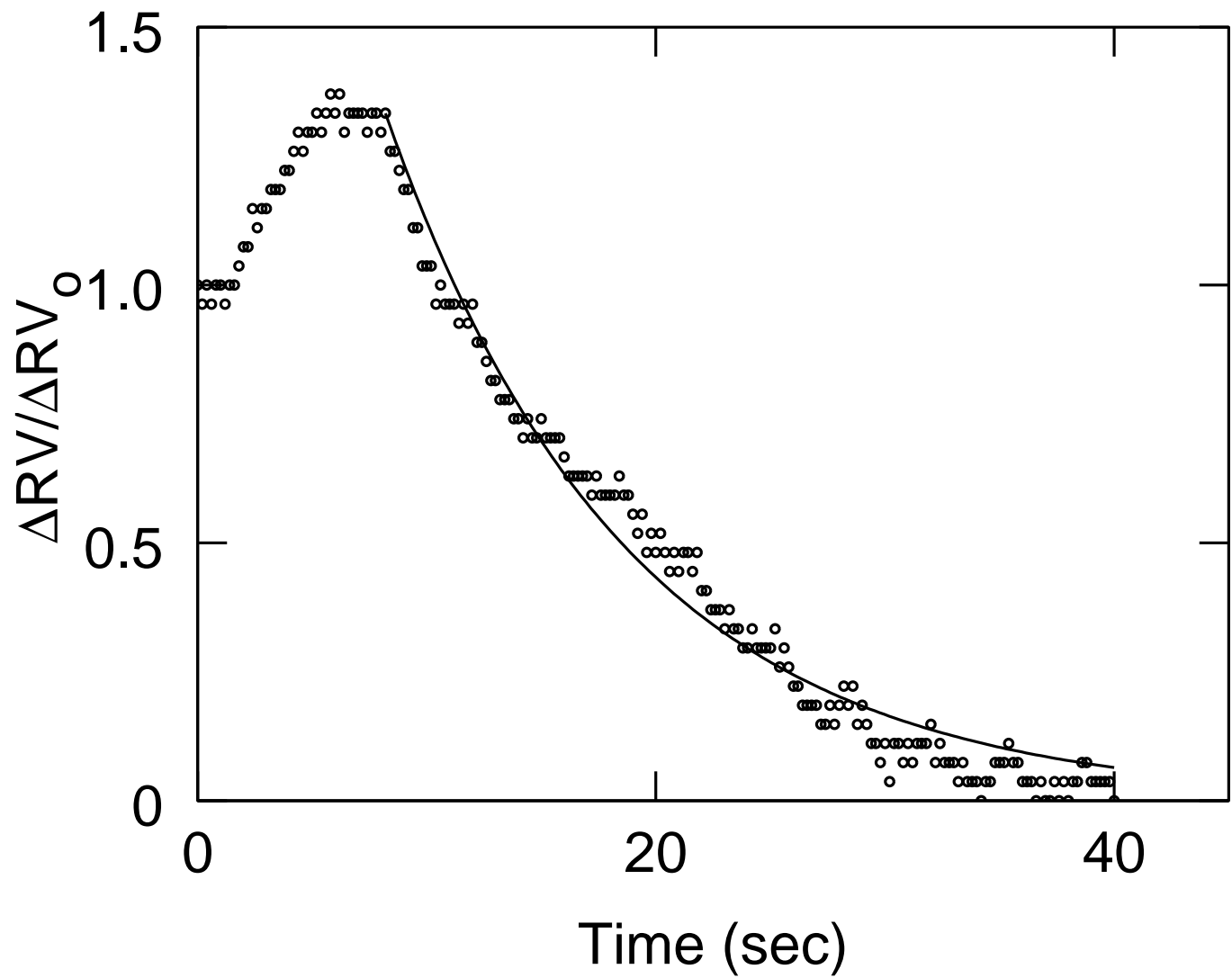
\end{document}